\documentclass[aps,prd,preprint,showpacs,nofootinbib]{revtex4}
\usepackage{latexsym,bm,amsmath,amssymb,graphicx,subfigure}
\DeclareMathAlphabet{\mathpzc}{OT1}{pzc}{m}{it}
\def\tr{\text{Tr}}
\begin{document}

\title{Lifshitz black holes in Einstein-Yang-Mills theory}
\author{Deniz Olgu Devecio\u{g}lu}
\email{dedeveci@metu.edu.tr}
\affiliation{Department of Physics, Faculty of Arts and  Sciences,\\
             Middle East Technical University, 06800, Ankara, Turkey}

\date{\today}

\begin{abstract}
We find that the four dimensional cosmological Einstein-Yang-Mills theory with $SU(2)$ gauge group admits 
Lifshitz spacetime as a base solution for the dynamical exponent $z>1$. Motivated by this, we next demonstrate numerically that the field 
equations admit black hole solutions which behave regularly on the horizon and at spatial infinity for different horizon topologies. 
The solutions depend on one parameter, the strength of the gauge field at the horizon, which is fine-tuned to capture the Lifshitz asymptotics at infinity. 
We also discuss the behavior of solutions and the change in Hawking temperature for black holes that are large or small with respect to the length 
scale $L$, which is itself fixed by the value of the cosmological constant.
\end{abstract}
 \maketitle
\section{Introduction}
The AdS/CFT conjecture has been a strong and versatile tool in the arsenal of high energy theory. The aptly named duality, 
relating conformal field theories to gravity in higher dimensions, has proven to be a powerful theoretical toolkit and provided great insight in
 high energy physics. Recently there has been a serious effort to trickle down to the energy scale of condensed matter and make holography accessible 
to strongly coupled systems which can be realized in experiments \cite{Hartnoll:2009sz,Balasubramanian:2008dm,Son:2008ye,Adams:2008wt} (and references 
therein). One of the approaches to achieve such duality is to impose an anistropic 
scaling symmetry on the boundary field theory
\begin{align}
 t\rightarrow \lambda^{z}t, \quad \vec x\rightarrow \lambda\vec x,\quad r\rightarrow \dfrac{r}{\lambda},
\end{align}
where $z$ is called the dynamical exponent. The symmetry algebra of field theories is controlled by $z$, e.g. $z=1$ generates the Poincar\'{e} 
group with special conformal symmetries, and when $z>1$ one ends up with different scalings for time and space which leads to non-relativistic 
field theories with Lifshitz symmetries; our main focus in this work. The bulk metric conjuring up these symmetries is found to be  
\begin{align}
 ds^{2}=L^{2}\Big(-r^{2z} dt^{2}+\dfrac{dr^{2}}{r^{2}}+r^{2}d\vec{x}^{2}\Big)\label{backmet},
\end{align}
with peculiar properties regarding causal structure and geodesics \cite{Kachru:2008yh,Hartnoll:2009sz}. Einstein gravity with a negative cosmological 
constant does not admit this type of anisotropic backgrounds as a solution. One should either consider higher derivative theories or matter couplings to 
source the metric. Once we depart from the Einstein gravity and add the higher curvature corrections, the amended theories begin to accommodate (\ref{backmet}) 
as a solution \cite{AyonBeato:2010tm}. On the other hand, the anisotropic backgrounds engineered with various types of matter Lagrangians 
\cite{Taylor:2008tg}, e.g. string theory motivated $p$-form fields \cite{Kachru:2008yh}, massive gauge fields, $U(1)$ fields with dilatonic-like 
couplings \cite{Tarrio:2011de} are better studied models for gravity duals. One of the first examples is the theory considered in \cite{Kachru:2008yh},
which is conjectured to be the gravitational dual of $2+1$ dimensional field theories modeling quantum critical behavior in strongly 
correlated electron systems. 

In principle, black hole solutions describe the finite temperature behavior of those dual non-relativistic field theories, which
renders them important objects in holography. Curvature corrections open up the way for large families of analytic black holes 
in different dimensions both for static and stationary Lifshitz spacetimes \cite{AyonBeato:2010tm,AyonBeato:2009nh,Cai:2009ac,Sarioglu:2011vz}. 
However, analytic black holes with matter fields for generic $z$ are rather rare \cite{Taylor:2008tg,Tarrio:2011de}. For a fixed value of $z$, 
several exact solutions were found \cite{Brynjolfsson:2009ct,Balasubramanian:2009rx,Pang:2009pd,Bertoldi:2009vn}. On the other hand, different types of 
numerical solutions were explored with generic $z$ values and for different horizon topologies \cite{Danielsson:2009gi,Dehghani:2010kd,Brenna:2011gp,Dehghani:2011tx,Mann:2009yx} 
for theories with massive gauge fields and $p$-forms. 

The matter Lagrangians with non-abelian gauge fields have been used in holographic superconductor models \cite{Gubser:2008wv,Gubser:2008zu}, 
with AdS/Schwarzschild black hole backgrounds. Recently the effects of Lifshitz scaling on these models have also been considered \cite{Lu:2013tza}. 
In this work we will first focus on a different and a simpler question: whether it is possible at all to support Lifshitz spacetime (\ref{backmet}) 
with non-abelian matter sources. To our knowledge, this has not been addressed previously elsewhere. Having answered the first in the affirmative, the 
second task we undertake is the dressing up of this background solution with black holes. There is a substantial literature on Einstein-Yang-Mills  
particle-like and black hole solutions \cite{Volkov:1998cc,Breitenlohner:1993es,Bartnik:1988am,Bizon:1990sr,Winstanley:1998sn} both in asymptotically flat and AdS backgrounds with different 
characteristics. For example, asymptotically flat, colored black holes \cite{Bizon:1990sr} admit finite range field strength, i.e. there is no global magnetic $SU(2)$ charge that 
makes them indistinguishable from Schwarzschild at infinity, whereas asymptotically AdS ones can possess global $SU(2)$ magnetic charge \cite{Winstanley:1998sn}. 
As we will show in what follows, Lifshitz asymptotics are quite different: Fields extend to infinity not only to endow black holes with $SU(2)$ charge but also to support 
Lifshitz spacetime. By abandoning asymptotic flatness, black holes with non-spherical horizon topologies can be constructed. 
Accordingly, we will consider three types of event horizon topologies, viz. planar, spherical and hyperbolic, with different gauge field ans\"{a}tze 
respecting the corresponding symmetries. For large black holes, these three types have similar behavior but differ significantly in the case of small event horizon radius. Our focus will be on the numerical evidence for 
the asymptotically Lifshitz black holes in cosmological EYM theory. We will not discuss the relation to the holographic dual field theories, 
which merits a separate significant problem on its own.

The outline of the paper is as follows: In Section \ref{asymptotics} we start with the equations of motion for the EYM system, 
state the ansatz for the planar symmetric YM fields and obtain the solution for the background metric (\ref{backmet}). We then 
set the stage for black hole solutions by dressing up the background metric and gauge fields with suitable functions in Section \ref{fieldeqn}. 
The subsections \ref{largerad} and \ref{seriesevent} are devoted to the series solutions of black holes at infinity and at the horizon, respectively. 
We next study the numerical black hole solutions of the theory in subsection \ref{numerics}. In Section \ref{temp} the Hawking temperature of the solutions 
we have found are analyzed. 
Finally we conclude with Section \ref{conc}.
\section{Lifshitz Asymptotics and $SU(2)$ Gauge Fields}{\label{asymptotics}}
The gravity theory we consider is the four dimensional cosmological EYM theory for the gauge group $SU(2)$ described by the action
\begin{align}
 S=\int d^{4}x\,\sqrt{-g}\Big((R-2\Lambda)-\dfrac{1}{2g_{\text{\tiny YM}}^{2}} \text{Tr}{\,F_{\mu\nu}F^{\mu\nu}}\Big)\label{action},\,
\end{align}
where $\Lambda$ is the cosmological and $g_{\text{\tiny YM}}^{2}$ is the gauge coupling constant in dimensions of 
$1/\text{length}^{2}$.\footnote{Here $ F_{\mu\nu}$ is the gauge field strength 
$F_{\mu\nu}\equiv F_{\mu\nu}^{a}T_{a}=\partial_{\mu}A_{\nu}-\partial_{\nu}A_{\mu}-i[A_{\mu},A_{\nu}]$ and we choose generators 
$T_{a}\equiv \tau_{a}/2,\,\, a=1,2,3$ with $\tau_{a}$ denoting Pauli matrices. The commutation relations and the normalization of generators are 
given as $ [T_{a},T_{b}]=i\epsilon_{abc}T_{c}$ and $ \tr\,T_{a}T_{b}=\delta_{ab}/2$, respectively. Throughout we use the conventions 
in which the signature of metrics is $(-,+,+,+)$, the Riemann tensor is taken as 
$R^{\mu}\,_{\nu\alpha\beta} = \partial_{\alpha} \Gamma^{\mu}\,_{\beta\nu}-\cdots$ and
$R_{\mu\nu}=R^{\alpha}\,_{\mu\alpha\nu}$.} 
In order to support backgrounds with anisotropic scaling symmetry, a naive approach is to make the coupling constants depend on the geometry, i.e. 
the parameter $z$. It is worth emphasizing that the path taken here is different from \cite{Winstanley:1998sn,VanderBij:2001ia}, 
in which AdS is already a vacuum for the gravitational sector and YM field is used only as a hair parameter, not for supporting the AdS geometry. 
In this work YM field will be used to source the metric (\ref{backmet}), so it has to decouple at $z=1$. Because of this major 
difference, we will not be able to recover the results of \cite{Winstanley:1998sn} in the conformal limit $z=1$. As we will show 
in the discussion below, ours is still an appropriate way to proceed.

Einstein field equations following from the action (\ref{action}) read
\begin{align}
R_{\mu\nu}-\Lambda g_{\mu\nu}=\dfrac{1}{g_{\text{\tiny YM}}^{2}}T_{\mu\nu},\label{einstein}
\end{align}
with the traceless YM stress-energy tensor defined as
\begin{align}
 T_{\mu\nu}\equiv\tr\, (F_{\mu}\,^{\alpha}F_{\nu\alpha}-\dfrac{1}{4}g_{\mu\nu}F_{\alpha\beta}F^{\alpha\beta}),\label{stress}
\end{align}
and the YM field equations
\begin{align}
 D_{\mu}F^{\mu\nu}=0,\label{ymeqns}
\end{align}
where the gauge covariant derivative is defined as $D_{\mu}\equiv\nabla_{\mu}-i[A_{\mu},\quad]$. 

The traceless nature of the stress-energy tensor allows us to determine the value of the cosmological constant from Einstein field equations. The trace
of (\ref{einstein}) when used with the metric (\ref{backmet}) yields
\begin{align}
\Lambda=-\dfrac{3+2z+z^2}{2L^2}.\label{cosmconst}
\end{align}
The next step is to consider the non-abelian gauge field configuration respecting the symmetry of the plane, which is a subgroup of the 
Poincar\'{e} group and studied extensively in \cite{Basler:1986yr,Basler:1984hw}. Additionally, we shall also restrict ourselves to the static 
and purely magnetic case. This restriction leads to the $SU(2)$ gauge connection 
\begin{align}
 A_{\mu}dx^{\mu}=w(r)T^{1}dx_{1}+w(r)T^{2}dx_{2}.\label{planeansatz}
\end{align}
For our purposes it is convenient to express the metric (\ref{backmet}) in a form which is analogous to the one that is 
commonly used\footnote{Here we are considering the planar case, whereas in the literature the spatial part of (\ref{backmetver2}) is typically 
spherical, with a different gauge field ansatz. The other cases can also be treated in a similar manner, which will be discussed later in 
the next section.} 
\cite{Volkov:1998cc,Breitenlohner:1993es,Winstanley:1998sn}
\begin{align}
 ds^2=L^{2}\Big(-S(r)^2 \mu(r) dt^2+ \dfrac{dr^2}{\mu(r)} + r^2 d\vec x^{2}\Big).\label{backmetver2}
\end{align}
Taking (\ref{planeansatz}), (\ref{backmetver2}) into account, the field equations (\ref{einstein}), (\ref{ymeqns}) reduce to the system
\begin{align}
 S^{-1}S'&=\dfrac{1}{2 L^{2}g_{\text{\tiny YM}}^{2}}\frac{(w')^{2}}{r},\label{eqm1}\\
 (\mu w')'&=\dfrac{w^{3}}{r^{2}}-\dfrac{1}{2 L^{2} g_{\text{\tiny YM}}^{2}}\dfrac{\mu (w')^{3}}{r},\label{eqm2}\\ 
 r \mu' + \mu + L^{2}r^{2}\Lambda &=-\dfrac{1}{2g_{\text{\tiny YM}}^{2}L^{2}}\Big(\dfrac{w^{4}}{2r^{2}}+\mu (w')^{2}\Big),\label{eqm3}
\end{align}
with prime denoting the ordinary derivative with respect to $r$.

Plugging in $S(r)= r^{z-1},\,\mu(r)= r^2$ and using (\ref{cosmconst}), it is straightforward to show that the Lifshitz spacetime 
(\ref{backmet}) is a solution for all $z > 1$ provided that the gauge field and the coupling constant are chosen as
\begin{align}
 w(r)=\pm\sqrt{z+1}\,r,\quad g_{\text{\tiny YM}}^{2}=\frac{1}{2L^{2}}\dfrac{(z+1)}{(z-1)}.\label{backsol}
\end{align}
The sign ambiguity of the gauge field can be deduced from the invariance of the field equations (\ref{eqm1}), (\ref{eqm2}), (\ref{eqm3}) under
$w(r)\rightarrow -w(r)$, which corresponds to a gauge transformation \cite{Volkov:1998cc}. Hence, in what follows we will proceed with the
positive sign gauge field. The solution we have found is basically a ``colorful plane with Lifshitz asymptotics''. Note also that $z>1$ in order to 
have real gauge fields, which signals the ``critical slowing down" of the possible dual field theories \cite{Kachru:2008yh}.

The conformal limit $z\rightarrow 1$ of (\ref{backsol}) is also peculiar. The YM part decouples from the gravity action and,  as well-known, 
the AdS spacetime is a solution of (\ref{einstein}) without matter fields, provided $\Lambda=-3/L^{2}$. Moreover, the decoupled gauge field is a 
solution to the pure YM part, which is in some sense the AdS analogue of the flatspace solution given in \cite{Basler:1986yr,Basler:1984hw}.

Having determined that the non-abelian YM matter is suitable for Lifshitz asymptotics, we can now continue and dress up this 
background geometry to obtain black hole solutions.  
\section{Field Equations}{\label{fieldeqn}}
In this section we first extend the metric and the gauge field ansatz to cover the other types of event horizon topologies, then cast the field 
equations in a way that is convenient for capturing the Lifshitz asymptotics for both the metric and the gauge field at large spatial distance. 

We will control the spatial part of the metric by introducing a parameter $k$ 
\begin{align}
 ds^{2}=L^{2}\Big(-S(r)^{2}\mu(r)dt^{2}+\dfrac{dr^{2}}{\mu(r)}+r^{2}d\Omega^{2}_{k}\Big)\label{fullmet},
\end{align}
where 
\begin{align}
d\Omega_k^2 \equiv \left\{\begin{array}{ll} d\theta^2 + \sin^2\theta d\phi^2 &;k=+1\\ 
d\theta^2 + \sinh^2\theta d\phi^2 &;k=-1 \\
d\theta^2 + d\phi^2 &;k=0 
\end{array} \right. \label{backmetsym}.
\end{align}
It is clear from this definition that, $k=0$ corresponds to the planar symmetric case we have discussed previously, $k=1$ yields the spherically symmetric 
metric, and $k=-1$ option is invariant under hyperbolic rotations.

The gauge field ansatz will change accordingly by taking into account the symmetries of the metric (\ref{backmetsym}).
The method for constructing symmetric gauge fields is developed in \cite{Forgacs:1979zs}. In addition to the $E(2)$ symmetric gauge field,   
we will also consider the following static $SU(2)$ connections that are invariant under $SO(3)$ and the connected part of $SO(2,1)$ 
\cite{Witten:1976ck,Forgacs:1979zs} 
\begin{align}
 A&=q(r) T^{3} dt+p(r) T^{3} dr +\big( w(r) T^{1}+u(r) T^{2}\big)d\theta \nonumber\\&+\big(w(r)\Omega_{k}(\theta)\,
 T^{2}-u(r)\Omega_{k}(\theta)\, T^{1}+\tilde\Omega_{k}(\theta) T^{3}\big) d\phi, \label{ansatzk1k0}
\end{align}
for $k=1,-1$, where $\Omega_{1}(\theta)\equiv\sin\theta,\,\Omega_{-1}(\theta)\equiv\sinh\theta,\,\tilde\Omega_{1}(\theta)\equiv\cos\theta,\,\tilde\Omega_{-1}(\theta)
\equiv\cosh\theta$.

This expression still has a $U(1)$ gauge freedom \cite{Breitenlohner:1993es}, which can be used to set $u(r)=0$. Next, with the help of the field equations, 
we see that $p(r)=0$ provided $w(r)\neq 0$. In order to simplify the discussion, we will only consider the gauge field strengths with vanishing electric 
part, i.e. $q(r)=0$. In fact this choice is rather restrictive. It was shown in 
\cite{Galtsov:1989ip} that, with appropriate asymptotics, the Reissner-Nordstr{\"o}m solution is the only static black hole with non-zero YM electric field. 
However, all of this was for asymptotically flat backgrounds and, obviously these arguments do not necessarily apply for Lifshitz spacetimes. Nevertheless, 
we shall restrict ourselves to the purely magnetic ansatz in this work.

Taking these considerations into account, we are thus led to 
\begin{align}
 A = \left\{\begin{array}{ll} w(r) T^{1} d\theta+( w(r)\Omega_{k}(\theta)\, T^{2}+ \tilde\Omega_{k}(\theta) \, T^{3})d\phi;\quad \text{for}\,\, k=\pm 1\\ 
w(r)T^{1}d\theta+w(r)T^{2}d\phi;\quad \text{for}\,\, k=0\\
\end{array} \right.\label{gaugef}.
\end{align}
Now utilizing the generalized metric (\ref{fullmet}) and the gauge field ansatz (\ref{gaugef}), the 
equations (\ref{eqm1}), (\ref{eqm2}), (\ref{eqm3}) can be cast into a general form covering all possible cases \cite{VanderBij:2001ia}
\begin{align}
 S^{-1}S'&=\dfrac{1}{2 L^{2}g_{\text{\tiny YM}}^{2}}\frac{(w')^{2}}{r},\label{eqm12}\\
 (\mu w')'&=\dfrac{w(w^{2}-k)}{r^{2}}-\dfrac{1}{2 L^{2} g_{\text{\tiny YM}}^{2}}\dfrac{\mu (w')^{3}}{r},\label{eqm22}\\ 
 r \mu' + \mu + L^{2}r^{2}\Lambda-k &=-\dfrac{1}{2g_{\text{\tiny YM}}^{2}L^{2}}\Big(\dfrac{(w^{2}-k)^{2}}{2r^{2}}+\mu (w')^{2}\Big).\label{eqm32}
\end{align}
Although this form of the field equations are helpful in exploring the constraints on the functions at the horizon, they are a bit impractical 
for numerical purposes. It is more appropriate to redefine the metric and the gauge field functions such that the Lifshitz vacuum 
(\ref{backmet}) can be explicitly recovered at large radius. One can achieve this with simple redefinitions
\begin{align}
 w(r)\equiv\sqrt{z+1}\,r h(r),\quad \mu(r)\equiv\dfrac{r^{2}}{g(r)^{2}},\quad S(r)\equiv r^{z-1}f(r)g(r), \quad w'(r)\equiv \sqrt{z+1}\, j(r).
\end{align}
It is obvious from these definitions that if all the unknown functions $f(r)$, $g(r)$, $h(r)$, $j(r)$ are unity in the large $r$ limit, 
i.e. when $r\gg1$, then we recover the Lifshitz background solution we have constructed for the EYM system with $k=0$.

All these assumptions, identifications and the coupling constants (\ref{cosmconst}), (\ref{backsol}) yield the following system of equations
\begin{align}
r f(r)^{\prime} & =-f(r)\Big((z-1)-\dfrac{j(r)^{2}}{2}(z-1)+\dfrac{g(r)^{2}h(r)^{4}}{4}(z^{2}-1)-\dfrac{g(r)^{2}}{4}(3+2z+z^{2})
+\dfrac{3}{2}\Big)\nonumber\\&-kf(r)g(r)^{2}\Big\{\dfrac{k}{4r^{4}}\dfrac{(z-1)}{(z+1)}-\dfrac{h(r)^{2}}{2r^{2}}(z-1)-\dfrac{1}{2r^{2}}\Big\},
\label{fulleqns1}\\
r j(r)^{\prime} & =  j(r)+g(r)^{2}h(r)^{3}(z+1)-
\dfrac{g(r)^{2}j(r)}{2}(z^{2}+2z+3)+\dfrac{g(r)^{2}h(r)^{4}j(r)}{2}(z^{2}-1)\nonumber\\&-k\Big\{g(r)^2\Big(\dfrac{h(r)^{2}j(r)}{r^{2}}(z-1)
-k \dfrac{j(r)}{2r^{4}}\dfrac{(z-1)}{(z+1)}+\dfrac{j(r)}{r^{2}}+\dfrac{h(r)}{r^{2}}\Big)\Big\},\label{fulleqns2}\\
r g(r)^{\prime} & =  \dfrac{g(r)j(r)^{2}}{2}(z-1)
+\dfrac{g(r)^{3}h(r)^{4}}{4}(z^{2}-1)-g(r)^{3}(3+2z+z^2)+\dfrac{3 g(r)}{2}\nonumber\\&+k g(r)^{3}\Big\{\dfrac{k}{4r^{4}}\dfrac{(z-1)}{(z+1)}-
\dfrac{h(r)^{2}}{2r^{2}}(z-1)-\dfrac{1}{2r^{2}}\Big\},\label{fulleqns3}\\
rh(r)^{\prime}&=j(r)-h(r).\label{fulleqns4}
\end{align}
Several observations are in order here. The highly nonlinear nature of the EYM system makes the analytic study difficult, and 
despite our efforts, we couldn't find an exact solution with non-trivial gauge field functions. Yet it is simple enough for working 
numerically, since we have reduced the system into a system of coupled first order ordinary differential equations 
with the functions having definite asymptotic values.

Secondly, terms explicitly involving $1/r^{2}$ and $1/r^{4}$ appear only in spherical and hyperbolic cases $k=\pm 1$. 
Exploiting this fact, we will assume that in the large $r$ limit, the spherical and hyperbolic spatial parts can be replaced in by a flat one 
\cite{Danielsson:2009gi}, \cite{Mann:2009yx}. Thus all of the unknown functions appearing in the numerical solutions will have the same asymptotic behavior, i.e. 
$f(r)=g(r)=h(r)=j(r)=1$.

Note that the three equations (\ref{fulleqns2}) to (\ref{fulleqns4}) form a closed system on their own, and equation (\ref{fulleqns1}) 
can be considered separately. In addition, the right hand side of (\ref{fulleqns1}) is linear in the function $f(r)$, which makes its 
normalization undetermined. This leads to a scaling of the initial value of $f$ at large $r$, which is essentially a gauge choice, i.e rescaling of the time coordinate 
\cite{Kachru:2008yh}. In order to get the correct asymptotics after the numerical integration, proper initial values must be chosen. 

There remains now to expand the functions $f(r),\,g(r),\,h(r),\,j(r)$ at large $r$ and separately at the horizon, for all possible values of 
the parameter $k$ but for a fixed value of $z$. One can extract a shooting parameter from the asymptotic form of the solutions to (\ref{fulleqns1}), 
(\ref{fulleqns2}), (\ref{fulleqns3}) and (\ref{fulleqns4}) provided there is one available with the given boundary conditions, and this is of 
paramount importance for the numerical study. 
\section{Series and Numerical Solutions\label{solutions}}
We now describe the results obtained by expanding the functions at large radius and at the horizon whose existence we assume presumably. 
The series solution will teach a great deal about the initial values and bounds on the functions defined in the previous section. We will then 
consider the numerical solutions of the system for various cases.
\subsection{Series solution for the large radius\label{largerad}}
First we look for the series solutions at large $r$, which in principle can confirm the plausibility of the assumption we have made in regards to the 
employment of the planar background for all horizon types at large $r$. The behavior of solutions is rather interesting for different values of $z$. 
It turns out that geometries with even integer dynamical exponent $z$ admit only planar solutions. However, all types of geometries are supported when $z$ is chosen to be an odd integer. 
In order to establish this result, we first fix the value of $z$ in equations (\ref{fulleqns1}), 
(\ref{fulleqns2}), (\ref{fulleqns3}), (\ref{fulleqns4}), then make a simple transformation $r=1/x$, and finally assume a power series expansion 
at small $x$ 
\begin{align}
 f(r)=\sum_{n=0}^{\infty}\tilde{f}_{n}x^{n},\quad g(r)=\sum_{n=0}^{\infty}\tilde{g}_{n}x^{n},\label{fexpand}
 \quad h(r)=\sum_{n=0}^{\infty}\tilde{h}_{n}x^{n},\quad j(r)=\sum_{n=0}^{\infty}\tilde{j}_{n}x^{n}
\end{align}
with the Lifshitz asymptotics, i.e. $\tilde{f}_{0}=\tilde{g}_{0}=\tilde{h}_{0}=\tilde{j}_{0}=1$. We insert these into the equations of motion 
(\ref{fulleqns1}), (\ref{fulleqns2}), (\ref{fulleqns3}), (\ref{fulleqns4}) and work order by order in $x$. We can summarize our findings as 
follows\footnote{To keep the following discussion simple, we only present our findings for the $z=2$ and $z=3$ cases. The generic behavior of the 
solutions are captured by the $z=2$ choice for even $z=4,6,8,\cdots$ or by the $z=3$ choice for odd $z=5,7,9,\cdots$.}:

For $z=2$ and $k=0$, we find
\begin{align}
 f(r)&=1-\frac{9 h_L}{2r^{4}} -\frac{1557}{176}\dfrac{h_L^2}{r^{8}}+\mathcal{O}(1/r^{16})+\cdots,\\
 g(r)&=1+\dfrac{6 h_L}{r^{4}}+ \frac{1143}{22}\dfrac{h_L^2}{r^{8}}+\mathcal{O}(1/r^{16})+\cdots,\\
 h(r)&=1+\dfrac{h_L}{r^{4}}+\frac{405}{44}\dfrac{h_L^2}{r^{8}}+\mathcal{O}(1/r^{16})+\cdots,\\
 j(r)&=1-\dfrac{3 h_L}{r^{4}}-\frac{2835}{44}\dfrac{h_L^2}{r^{8}}+\mathcal{O}(1/r^{16})+\cdots.
\end{align}
However, for $z=3$ and with generic $k$, we get
\begin{align}
  f(r)&=1+\frac{k}{2r^{2}}+\frac{127}{1352}\frac{k^2}{r^{4}}+\mathcal{O}(1/r^{5})+\cdots,\\
 g(r)&=1+\frac{23}{676}\frac{k^2}{r^{4}}+ \frac{12 h_L}{r^{5}}+\mathcal{O}(1/r^{6})+\cdots,\\
 h(r)&=1-\dfrac{3}{338}\frac{k^2}{r^{4}}+\dfrac{h_L}{r^{5}}+\mathcal{O}(1/r^{6})+\cdots,\\
 j(r)&=1+\dfrac{9}{338}\frac{k^2}{r^{4}}-\frac{4 h_L}{r^{5}}+\mathcal{O}(1/r^{6})+\cdots,
\end{align}
where we have only one arbitrary parameter $h_{L}$ characterizing both solutions at large $r$. Let us emphasize that the discrepancy between even and odd 
$z$ follows from the expansion (\ref{fexpand}) we have considered. There may be fractional powers of $x$ in the expansion (\ref{fexpand}) 
which can remedy the situation for the even $z$ case. It is also possible that we have made an inappropriate choice of coordinates to discuss the solutions for large $r$.
Nevertheless, we fix $z=3$ in the numerical part of the calculations (see section \ref{numerics}) for the sake of clarity.
\subsection{Series solution about the event horizon \label{seriesevent}}
Let us now focus on the series solution about the presumed horizon. In order to have a non-extremal black hole, $g_{tt}$ and $g_{rr}$ components of the metric 
(\ref{fullmet}) must have a simple zero and a simple pole \cite{Danielsson:2009gi}, \cite{Mann:2009yx} at the finite horizon $r=R_{0}$. This assumption 
leads to the following horizon expansions of the functions 
\begin{align}
f(r)&=\sqrt{r-R_{0}}\,\sum_{n=0}^{\infty}f_{n}(r-R_{0})^{n},\\
g(r)&=\dfrac{1}{\sqrt{r-R_{0}}}\,\sum_{n=0}^{\infty}g_{n}(r-R_{0})^{n}.
\end{align}
At this stage it is worthwhile to discuss the constraints on the gauge field functions at the horizon in order to construct the series expansion 
for the functions $h(r)$ and $j(r)$. These constraints can easily be seen from the general form of the field equations (\ref{eqm12}), (\ref{eqm22}), 
(\ref{eqm32}) we have discussed in section \ref{fieldeqn}. 
This set implies that the gauge field function $w(r)$ and its derivative must be related at the horizon as
\begin{align}
 w'(R_{0})=\dfrac{w(R_{0})(w^{2}(R_{0})-k)}{\Big(k R_{0}-\dfrac{1}{2 g_{\text{\tiny YM}}^{2} L^{2}}\dfrac{(w^{2}(R_{0})-k)^{2}}{2R_{0}}
 -L^{2}R_{0}^{3}\Lambda\Big)},
\end{align}
which amounts to relating the expansion coefficients on the horizon
\begin{align}
 j(R_{0})=j_{0}=\frac{ 2 h_0 R_{0} \left(h_0^2 R_{0}^2 (z+1)-k\right)}{2 k
   R_{0}+ R_{0}^3 \left(z^2+2 z+3\right)-\frac{(z-1) \left(k-h_0^2 R_{0}^2 (z+1)\right){}^2}{R_{0} (z+1)}};\quad \text{for}\quad z>1\label{jzero}
\end{align}
where $w(R_{0})=\sqrt{z+1}R_{0}h_{0}$ with the definition $h_{0}\equiv h(R_{0})$.
The subtle difference between the planar and the other cases shows itself here. When $k=0$, the horizon radius cancels out, and $j_{0}$ depends only 
on $h_{0}$ and the dynamical exponent $z$. To make the meaning of $h_{0}$ clear, consider a non-coordinate basis for the one-forms \cite{Kachru:2008yh}
\begin{align}
 \theta_{t}=Lr^{z}f(r) dt,\quad \theta_{x_{i}}=Lr dx^{i},\quad \theta_{r}=L\frac{g(r)}{r}dr,\quad i=1,2
\end{align}
in which the planar metric (\ref{fullmet}) takes the form $ds^{2}=\eta^{\mu\nu} d\theta_{\mu}d\theta_{\nu}$ with $\eta^{\mu\nu}=\text{diag}(-1,1,1,1)$. 
The gauge connection simply follows as
\begin{align}
 A=\frac{\sqrt{z+1}}{L}h(r)(T^{1}\theta_{1}+T^{2}\theta_{2}).
\end{align}
This suggests that $h_{0}$ can be considered as the strength of the gauge field at the horizon, up to some normalization.
There is also an upper bound for the gauge field function $w(r)$ for a given horizon radius $R_{0}$, 
which follows from the condition for a regular horizon, i.e
\begin{align}
 \dfrac{d\mu}{dr}\bigg|_{r=R_{0}}>0.
\end{align}
Then, with the help of (\ref{eqm32}), one finds that
\begin{align}
k -\dfrac{1}{2 g_{\text{\tiny YM}}^{2} L^{2}}\dfrac{(w^{2}(R_{0})-k)^{2}}{2R_{0}^{2}}
 -L^{2}R_{0}^{2}\Lambda>0. \label{50}
\end{align}
In terms of $w(R_{0})=\sqrt{z+1}\,h_{0}$, this inequality further simplifies to
\begin{align}
\dfrac{R_{0}^{2}(z+1)\Big(2k+R_{0}^{2}(3+2z+z^{2})\Big)}{(z-1)}> (k-R_{0}^{2}(z+1)h_{0}^{2})^{2}.\label{ineq}
\end{align}
The inequality (\ref{ineq}) is rather important for numerical purposes. It weakly constrains the strength of the gauge field at the horizon, which in turn 
reduces the possible values for the shooting parameter $h_{0}$. For $k=0$, $h_{0}$ 
is solely bounded by the $z$ value. There is no dependence on the horizon radius; i.e if a numerical solution is found for the system with a fixed 
value of $h_{0}$, then it will always remain to be a solution for different radii. On the other hand, for the other topologies $k=\pm 1$, the gauge field 
strength changes with the changing horizon radius. The hyperbolic case $k=-1$ demands special attention regarding the value of the event horizon radius. 
By virtue of (\ref{50}), there is a lower bound on the event horizon radius for fixed $z$
\begin{align}
 |\Lambda|> \dfrac{1}{L^{2} R_{0}^{2}}(1+\dfrac{1}{4 g_{\text{\tiny YM}}^{2} R_{0}^{2} L^{2}}).\label{hyperbound}
\end{align}
The bound and the relations above can also be extracted from near horizon expansions. Assuming that the functions $h(r),j(r)$ 
are finite on the horizon, they read 
\begin{align}
 h(r)&=\sum_{n=0}^{\infty}h_{n}(r-R_{0})^{n},\label{hexpand}\\
 j(r)&=\sum_{n=0}^{\infty}j_{n}(r-R_{0})^{n}.\label{jexpand}
\end{align}
Inserting the expansions (\ref{fexpand}), (\ref{hexpand}), (\ref{jexpand}) into (\ref{fulleqns1}), (\ref{fulleqns2}), (\ref{fulleqns3}), (\ref{fulleqns4}), 
one finds solutions depending on two free parameters $h_{0}$, the strength of the gauge field at the horizon, and $R_{0}$, the horizon radius for 
a fixed $z$ value.  

As a simple example, for $z=2$ and $k=0$, one gets
\begin{align}
 g_0&\to \frac{\sqrt{2R_{0}}}{\sqrt{11-3 h_0^4}},\\
 j_0&\to \frac{6 h_0^3}{11-3 h_0^4},\\
 h_1&\to \frac{h_0 \left(3 h_0^4+6 h_0^2-11\right)}{\left(11-3 h_0^4\right) R_{0}},\\
 g_1&\to \frac{\sqrt{2} \left(18 h_0^8+27 h_0^6-99 h_0^4+121\right)}{\left(11-3 h_0^4\right){}^{5/2}
   \sqrt{R_{0}}},\\
   f_1&\to \frac{f_0 \left(-27 h_0^8+9 h_0^6+165 h_0^4-242\right)}{\left(11-3 h_0^4\right){}^2 R_{0}}.\\ \ \nonumber 
\end{align}
Note that, all of the coefficients depend on two parameters $h_{0},R_{0}$. 
Although $f_{0}$ appears to be a free parameter, it is in fact just an overall normalization factor as noted earlier in the penultimate paragraph of 
section \ref{fieldeqn}. The bound on $h_{0}$ is now clear. In order to have real values for $g_{0}$, $h_{0}$ must be smaller than a value depending on $z$, 
and for $z=2$, $k=0$ the strength of the gauge field must be $h_{0}^{4}<11/3$ which is consistent with (\ref{ineq}). 
Finally the value of $j_{0}$ (\ref{jzero}) is also recaptured here.

To sum up, we have paved the way for numerical computation, by finding the initial values for functions in terms of $h_{0}$ and $R_{0}$. Now 
fixing one of the two parameters, namely the event horizon radius $R_{0}$, the shooting method can be used to search for numerical solutions. 
For a fixed value of $R_{0}$, we numerically evolve the functions and make them converge to unity at infinity by fine tuning the initial value 
$h_{0}$. The behavior of solutions differs considerably for small and large horizon radius values, and it also depends on the topology.
\subsection{Numerical solutions\label{numerics}}
We begin with the larger black holes, and fix $z=3$ in order to compare results for different values of $k$. It turns out that
there is a unique critical value of $h_{0}$ within the allowed region described by (\ref{ineq}), where we have the desired asymptotics. 
This is quite different from what was observed in asymptotically flat or AdS analogues of these black holes, where solutions are indexed by an integer 
$n$ that has the meaning of the node number for the gauge field amplitude $w_{n}(r)$ 
\cite{Bartnik:1988am,Bizon:1990sr,Breitenlohner:1993es,Winstanley:1998sn,Volkov:1998cc}. 

\begin{figure}[!htbp]
\centering
\includegraphics[width=0.8\textwidth]{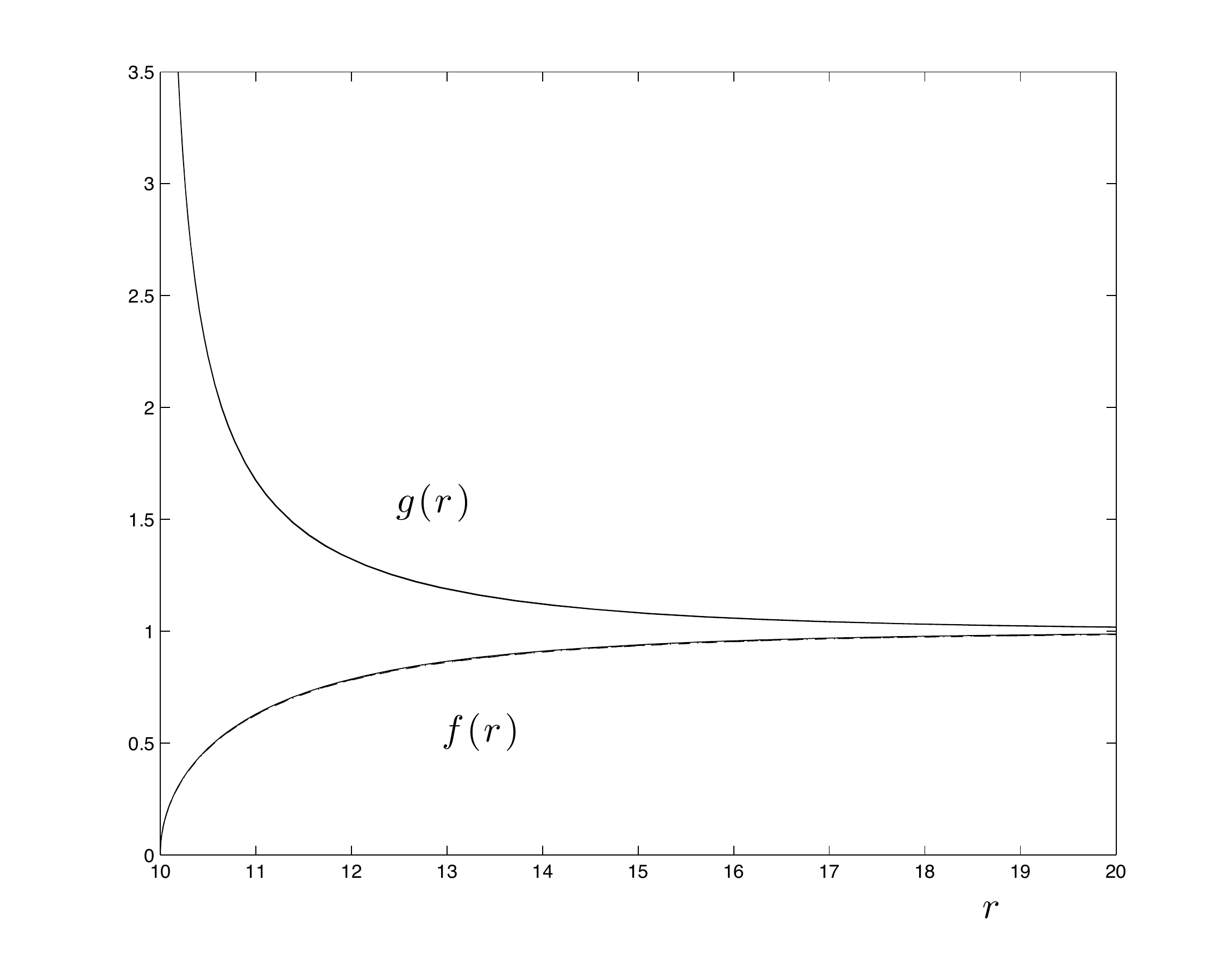}
\caption{The figure plots the metric functions $f(r)$ and $g(r)$ as a function of radius $r$. This is an example of a large black hole with 
$R_{0}=10$, where the plots overlap for all values of $k$.}\label{fig1}
\end{figure}

\begin{figure}[!htbp]
\centering
\includegraphics[width=0.8\textwidth]{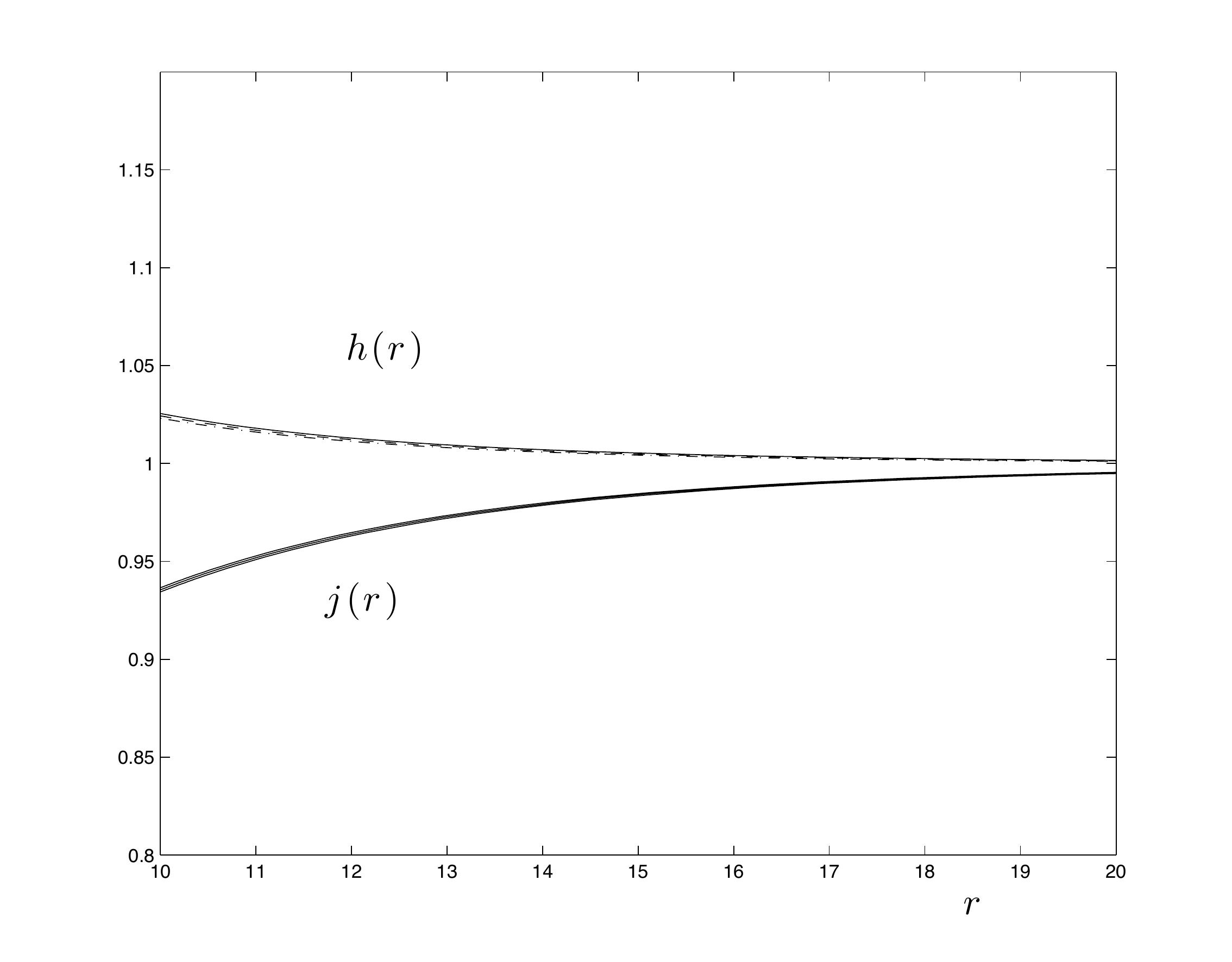}
\caption{The figure shows the gauge field functions $h(r)$ and $j(r)$ as a function of radius $r$ with $R_{0}=10$. The initial values of functions for different 
topologies are very close to each other. Graphs for different topologies merge into one.} \label{fig2}
\end{figure}

Setting $R_{0}=10$, we see from figures \ref{fig1},\ref{fig2} that, for large black holes the solutions behave similarly regardless of the topology of the 
event horizon. Although we plot the functions for all values of $k$, the graphs coalesce into one with a small difference between their 
shooting parameters $h_{0}$. The metric functions $f(r)$ and $g(r)$ start from zero and infinity, respectively, then converge to one monotonically. 
We have the following results for the initial value of the gauge field function, i.e. the shooting parameter
\begin{align}
 h_{0}= \left\{\begin{array}{ll} 1.025530137,\quad \text{for}\,\,& k=1,\\
  1.023139854,\quad \text{for}\,\,& k=-1,\\
  1.024335678,\quad \text{for}\,\,& k=0.\end{array}\right.
\end{align}
The value of $j_{0}$ simply follows from (\ref{jzero}).
\begin{figure}[!htbp]
\centering
\includegraphics[width=0.80\textwidth]{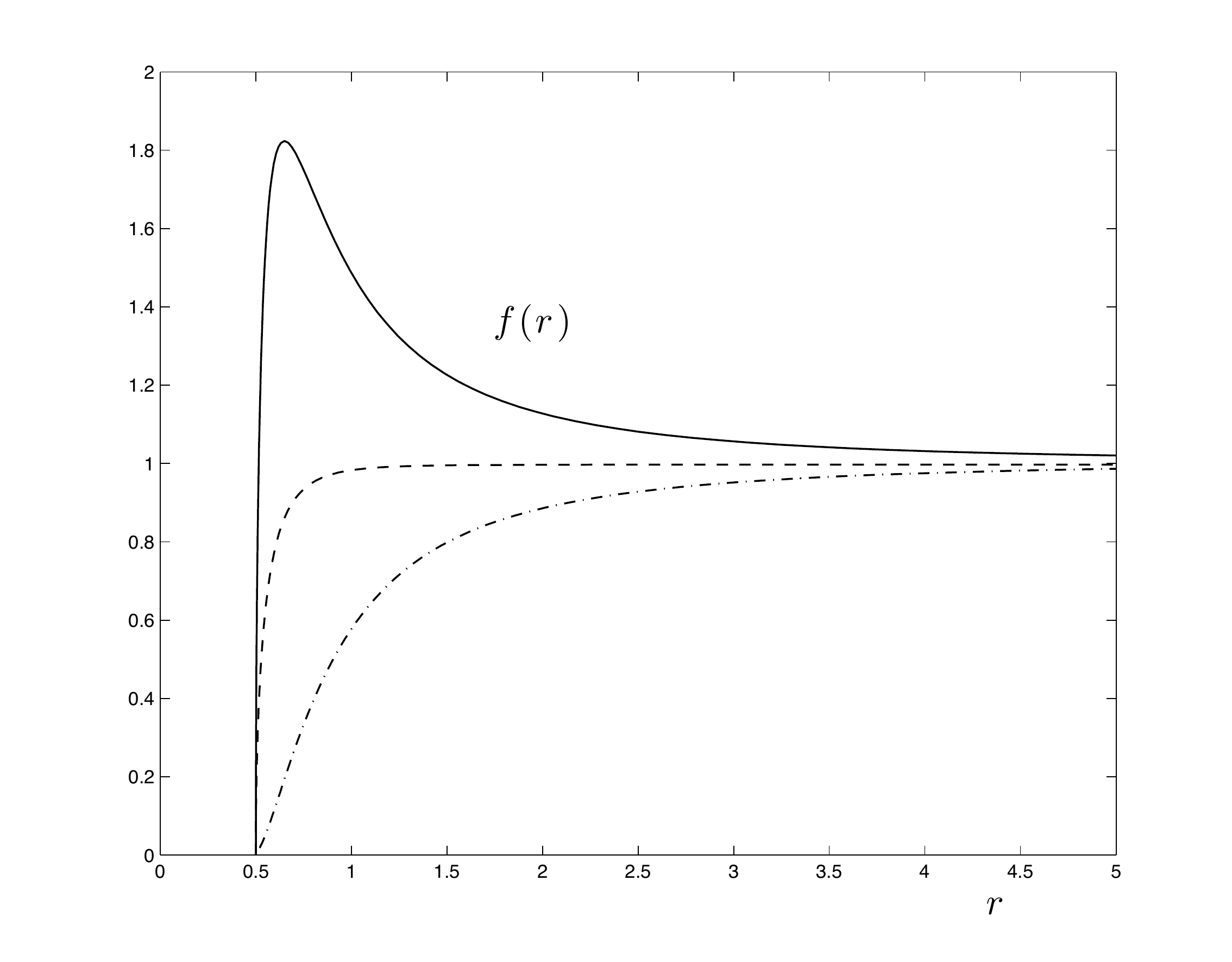}
\caption{A small black hole with $R_{0}=0.5$. Figure shows the metric function $f(r)$ for different cases $k=1,-1,0.$ The solid line corresponds to 
$k=1$, the dashed line to $k=0$ and dot-dashed line represents $k=-1$, respectively.} \label{fig3}
\end{figure}
\begin{figure}[!htbp]
\centering
\includegraphics[width=0.80\textwidth]{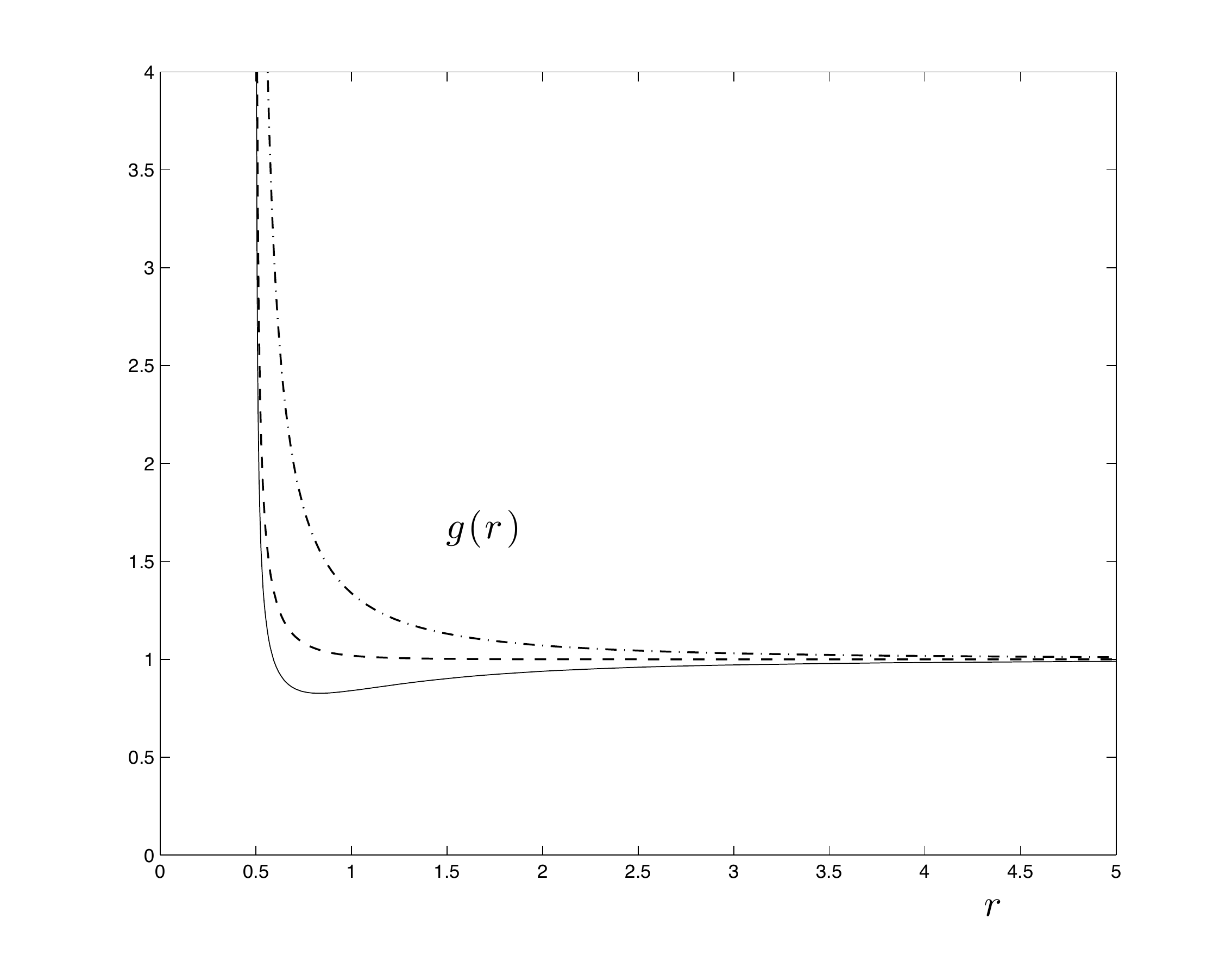}
\caption{The figure illustrates the metric function $g(r)$ with a small radius $R_{0}=0.5$. The solid line indicates $k=1$, while the $k=0$ and $k=1$ 
cases are represented by dashed and dot-dashed lines, respectively.} \label{fig4}
\end{figure}
\begin{figure}[!htbp]
\centering
\subfigure{\includegraphics[width=0.78\textwidth]{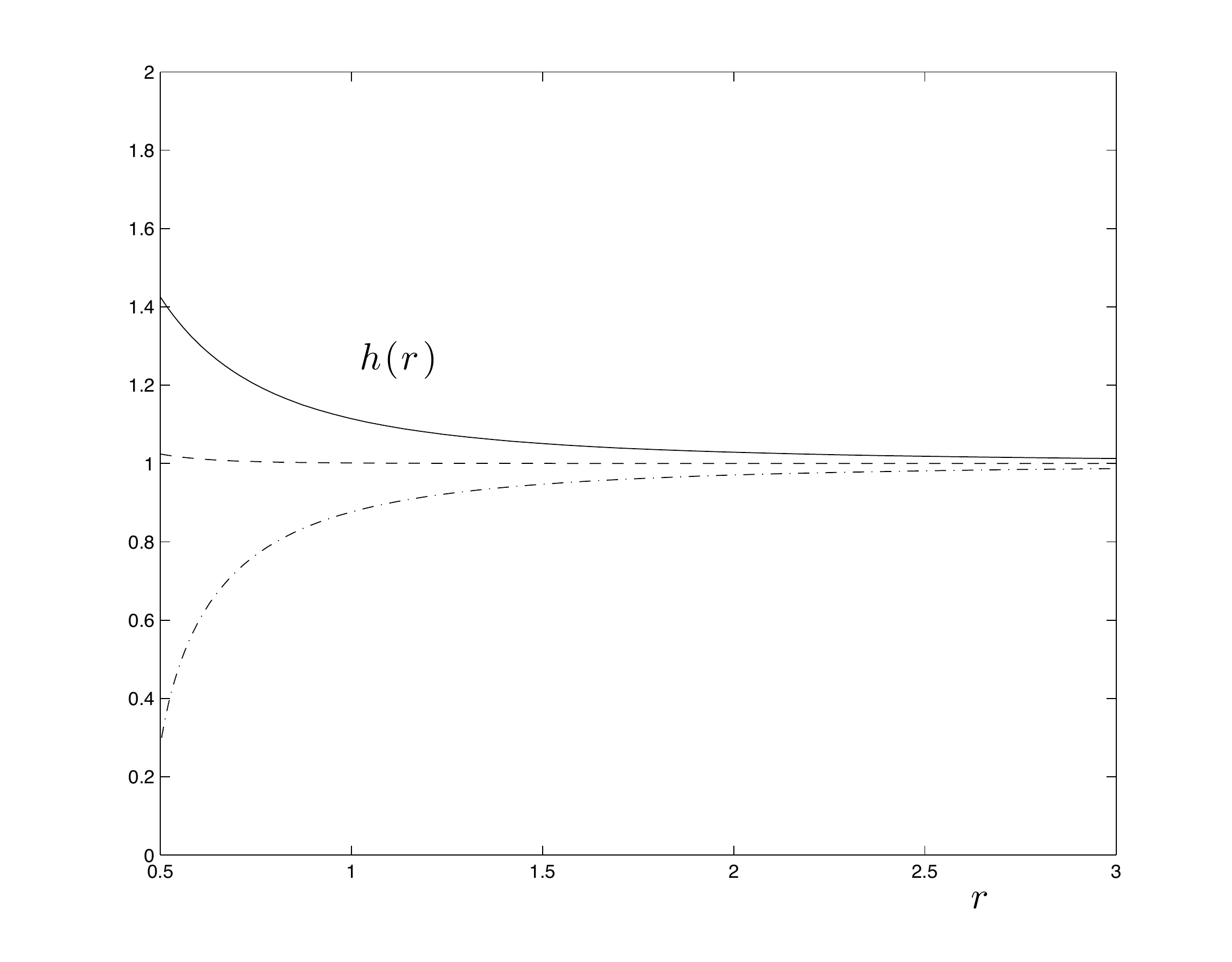}}\\
\subfigure{\includegraphics[width=0.78\textwidth]{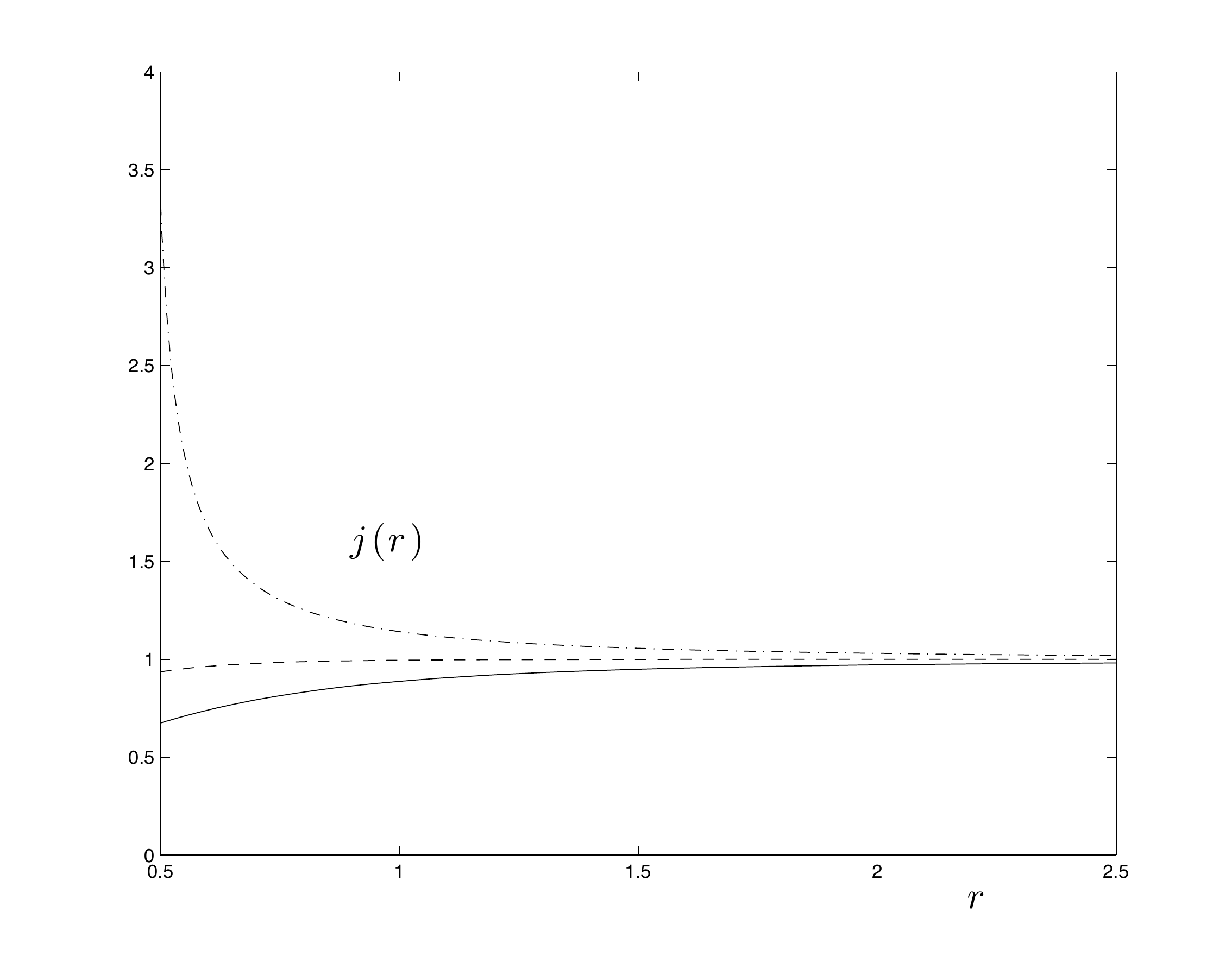}}
\caption{The gauge field function $h(r)$ is displayed on the top and $j(r)$ at the bottom, both as functions of $r$. In both 
graphs $R_{0}=0.5$. The solid line indicates $k=1$, while the $k=0$ and $k=1$ 
cases are represented by dashed and dot-dashed lines, respectively. } \label{fig5}
\end{figure}

\begin{figure}[!htbp]
\centering
\includegraphics[width=0.80\textwidth]{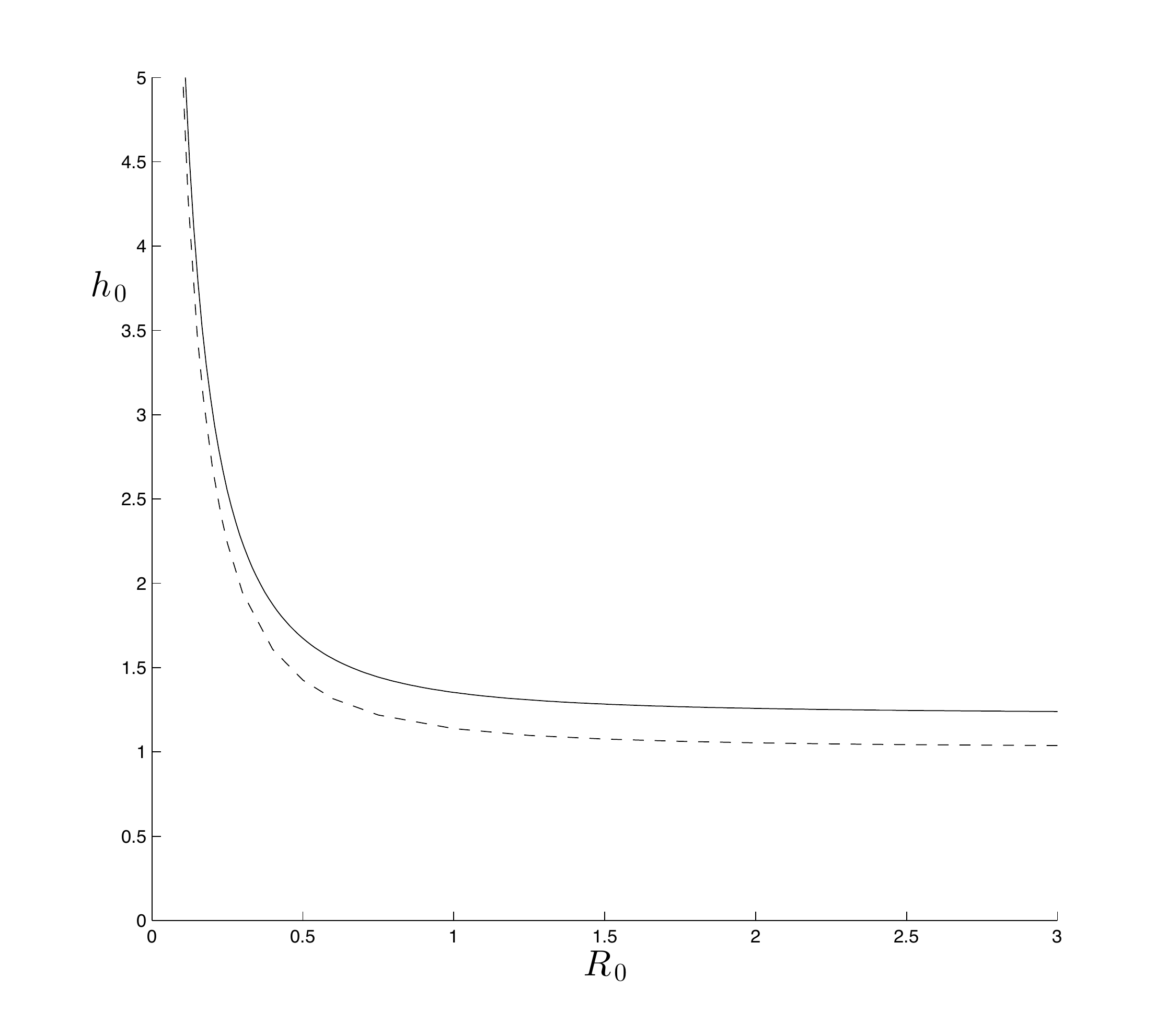}
\caption{The inequality (\ref{ineq}) as a function of $R_{0}$ is plotted with a solid line for $k=1$. The dashed line corresponds to the 
numerical values of $h_{0}$ as a function of $R_{0}$ for spherically symmetric black holes.} \label{fig6}
\end{figure}

\begin{figure}[!htbp]
\centering
\includegraphics[width=0.80\textwidth]{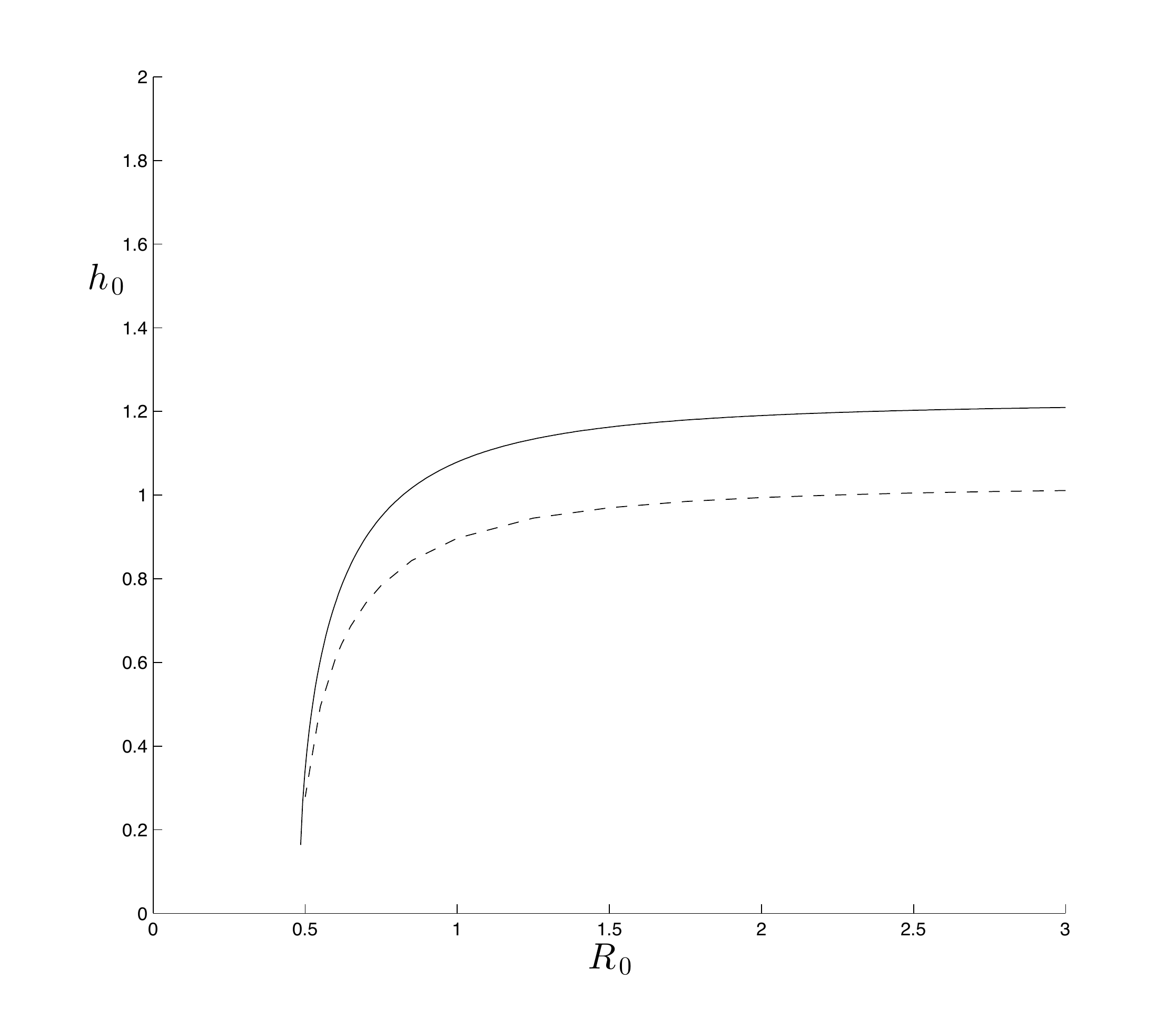}
\caption{The inequality (\ref{ineq}) as a function of $R_{0}$ is plotted with a solid line for $k=-1$. The dashed line corresponds to the 
numerical values of $h_{0}$ as a function of $R_{0}$ for hyperbolically symmetric black holes. The lower bound (\ref{hyperbound}) 
on the horizon radius is apparent.} \label{fig7}
\end{figure}

We then fix  $R_{0}=0.5$ in order to investigate the smaller black holes. The behavior of the solutions changes drastically. First of all, 
the functions of spherical and hyperbolic solutions decay appreciably slower, and moreover the shooting parameters i.e. $h_{0}$ differ considerably. 
From figures \ref{fig3} and \ref{fig4}, we see that for the spherical case the metric function $f(r)$ makes a peak first and then converges to unity, unlike the 
planar and hyperbolic cases where the functions monotonically converge to one. The other metric function $g(r)$ reaches a minimum then approaches 
to one for the spherically symmetric black holes. It turns out that for small black holes we have the following gauge field strengths (see figure \ref{fig5})
\begin{align}
 h_{0}= \left\{\begin{array}{ll} 1.425617169,\quad \text{for}\,\,& k=1,\\
  0.278652475,\quad \text{for}\,\,& k=-1,\\
  1.024335678,\quad \text{for}\,\,& k=0.\end{array}\right.
\end{align}
Having seen the differences between large and small black holes, let us now compare the analytic bound (\ref{ineq}) with the values of $h_{0}$ 
for different radii. For planar black holes, a unique value of $h_{0}$ is sufficient for all event horizon radii. Meanwhile, for the spherical 
case one needs larger gauge fields for small radii, and hyperbolic 
ones can support weaker gauge fields as the radius gets smaller. A similar behavior was observed for the abelian field strength in the works of 
\cite{Danielsson:2009gi,Mann:2009yx}. For clarity, we plot $h_{0}$ versus $R_{0}$ both for spherical (figure \ref{fig6}) and hyperbolic (figure \ref{fig7}) 
cases as well. The solid line depicts the solution of the inequality (\ref{ineq}) as a function of $R_{0}$ and the dashed line is the numerical 
values obtained from the shooting method. Evidently the bound 
(\ref{ineq}) is saturated as the horizon radius $R_{0}$ gets smaller. It is worth emphasizing that the lower limit on the horizon radius (\ref{hyperbound}) 
for $z=3$ is consistent with the numerical results, i.e. from the figure \ref{fig7} we see that there is no solution below $R_{0}\sim 0.48$. 
\section{Thermal behavior\label{temp}}
\begin{figure}[ht]
\centering
\includegraphics[width=0.65\textwidth]{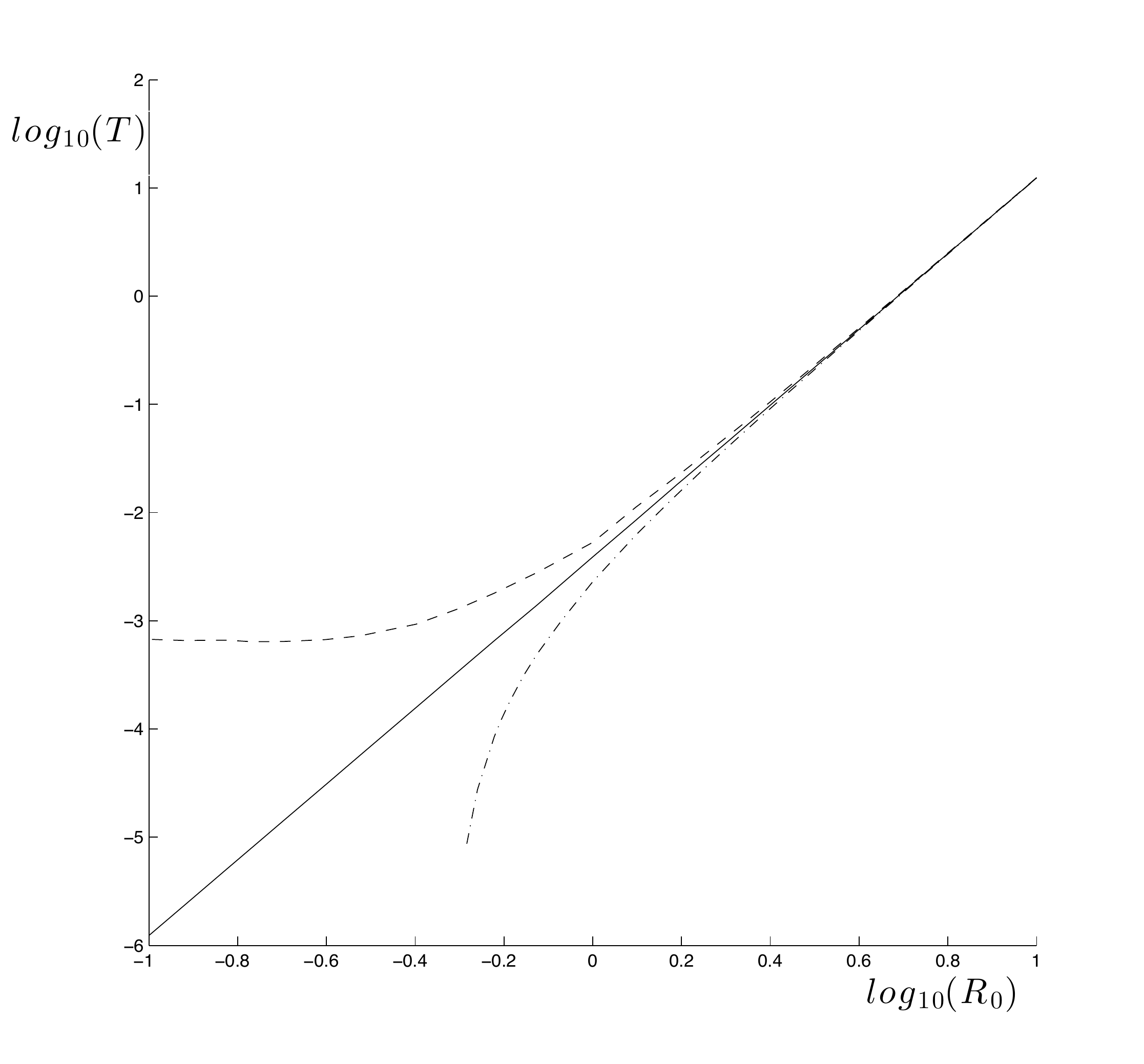}
\caption{Temperature versus horizon radius for $z=3$. The different topologies are represented by a solid line $k=0$, by a dashed line $k=1$ and a dot-dashed line 
$k=-1$.} \label{fig8}
\end{figure}

Finally let us compute the temperature and discuss the thermal behavior of these black holes. 
We resort to the Euclidean metric obtained by a Wick rotation to compute the temperature, which leads to the following expression \cite{Dehghani:2011tx}
\begin{align}
 T=\dfrac{f_{0} R_{0}^{z+1}}{4\pi g_{0}},
\end{align}
where $f_{0}$, $g_{0}$ are the expansion coefficients in the near horizon limit. The general expression from the series solution near the 
horizon determines $g_{0}$ in terms of $k$, $h_{0}$, $R_{0}$ and $z$:
\begin{align}
 g_0= \dfrac{\sqrt{2(z+1)} R_{0}^{3/2}}{(2 h_0^2 k R_{0}^2 (z-1)+h_0^4 R_{0}^2 (1-z) (z+1)-k^2
   \frac{(z-1)}{(z+1)}+2 k +R_{0}^2 (3+ 2z+z^{2}))^{1/2}}.
\end{align}
Recall that the coefficient $f_{0}$ is to be determined from the normalization of the numerical solution, so it depends on the shooting parameter $h_{0}$. 
Therefore, fixing $z=3$, the temperature now depends only on the horizon radius and the topology. After finding several numerical solutions for 
different $R_{0}$ values, we plot figure \ref{fig8} by computing the temperature within the limits of numerical accuracy. 
It is clear from this figure that as the radius gets smaller, black holes get cooler 
with different rates. Hyperbolic ones have a higher cooling rate then the planar ones, and the spherical black holes are hotter for small radius. 
In the large $R_{0}$ limit, the temperatures become identical just like the solutions. The thermal behavior of these black holes is opposite 
to their AdS counterparts, where the Hawking temperature increases with the ever decreasing radius causing thermal instability. 
Moreover, it is clear that the EYM black holes do not exhibit Hawking-Page transition. A similar thermal behavior is observed for the Lifshitz black 
holes supported by abelian $p$-forms \cite{Danielsson:2009gi,Mann:2009yx} which indicates that the black holes become extremal i.e. they have zero Hawking 
temperature in the vanishing black hole size.

\section{Conclusions\label{conc}}
In this work, we have studied the Lifshitz black holes with different horizon topologies in four dimensional cosmological EYM theory. 
After obtaining the gauge field that supports the Lifshitz spacetime (\ref{backmet}), we have found numerical black hole solutions with 
different horizon topologies by suitably fine-tuning the gauge field strength at the horizon. Through the series solution of the field equations, 
we have found a quite interesting property: The geometries with odd $z$ support black holes with different horizon topologies, whereas 
for even $z$ only planar ones are supported. Thus we have fixed $z=3$ in order to investigate all possible scenarios. From numerical results, 
we have observed that the behavior of solutions for different topologies changes considerably for small black holes whereas it becomes identical 
for large horizon black holes. We have also analyzed the thermal behavior of the numerical solutions by computing the Hawking temperature for all types of 
black holes. We have found that there is a rapid decay in temperature as the black hole radius gets smaller, and moreover black holes do not display 
Hawking-Page transition. In this respect, the EYM black holes and the abelian counterparts \cite{Danielsson:2009gi,Mann:2009yx} 
have quite similar characteristics, but they both differ considerably from their conformal cousins and some of the Lifshitz black hole solutions 
to string theory \cite{Amado:2011nd}.

One of the most important questions to ask is the use of EYM theory in non-relativistic holography. Certainly, Lifshitz spacetimes and black holes 
with non-abelian matter sources deserve further attention. Although the holographic description of 
matter Lagrangians with abelian and scalar fields are studied up to some extent, there is not much work done on EYM theory in which these solutions 
can find a practical application.

A further direction of research would be to consider the extension of these black holes. First, the existence of analogous solutions 
can be considered by extending the $SU(2)$ symmetry ansatz to higher spacetime dimensions. It would also be interesting to investigate the generalization of the gauge group $SU(2)$ 
to $SU(N)$. In another vein, here we have only considered a purely magnetic part; One could still extend this ansatz by turning on the function 
$q(r)$ in (\ref{ansatzk1k0}) and look for the existence of dyonic black holes. It would certainly be of interest if the non-abelian 
counterparts of Lifshitz solitons \cite{Danielsson:2009gi,Mann:2009yx,Mann:2011bt} could be found.
\begin{acknowledgments}
 I thank {\"O}zg{\"u}r Sar{\i}o\u{g}lu for suggesting this problem, his valuable comments and critical reading of the manuscript. 
 I also thank Bayram Tekin, Dieter Van den Bleeken and Robert Mann for their comments, suggestions, and G{\"o}khan Alka\c{c} for his help 
 in the numerical part of the calculations. This work is partially supported by the Scientific and Technological Research 
 Council of Turkey (T{\"U}B\.{I}TAK) Grant No.113F034.
\end{acknowledgments}


\begin{thebibliography}{99}
\bibitem{Balasubramanian:2008dm} 
  K.~Balasubramanian and J.~McGreevy,
  Phys.\ Rev.\ Lett.\  {\bf 101}, 061601 (2008)
  [arXiv:0804.4053 [hep-th]].
 
\bibitem{Hartnoll:2009sz} 
  S.~A.~Hartnoll,
  Class.\ Quant.\ Grav.\  {\bf 26}, 224002 (2009)
  [arXiv:0903.3246 [hep-th]].
\bibitem{Son:2008ye} 
  D.~T.~Son,
  Phys.\ Rev.\ D {\bf 78}, 046003 (2008)
  [arXiv:0804.3972 [hep-th]].
  
\bibitem{Adams:2008wt} 
  A.~Adams, K.~Balasubramanian and J.~McGreevy,
  JHEP {\bf 0811}, 059 (2008)
  [arXiv:0807.1111 [hep-th]].
\bibitem{Kachru:2008yh} 
  S.~Kachru, X.~Liu and M.~Mulligan,
  Phys.\ Rev.\ D {\bf 78}, 106005 (2008)
  [arXiv:0808.1725 [hep-th]].

\bibitem{AyonBeato:2010tm} 
  E.~Ayon-Beato, A.~Garbarz, G.~Giribet and M.~Hassaine,
  JHEP {\bf 1004}, 030 (2010)
  [arXiv:1001.2361 [hep-th]].
\bibitem{Taylor:2008tg} 
  M.~Taylor,
  ``Non-relativistic holography,''
  arXiv:0812.0530 [hep-th].
 
\bibitem{Tarrio:2011de} 
  J.~Tarrio and S.~Vandoren,
  JHEP {\bf 1109}, 017 (2011)
  [arXiv:1105.6335 [hep-th]].

\bibitem{Cai:2009ac} 
  R.~-G.~Cai, Y.~Liu and Y.~-W.~Sun,
  JHEP {\bf 0910}, 080 (2009)
  [arXiv:0909.2807 [hep-th]].
\bibitem{AyonBeato:2009nh} 
  E.~Ayon-Beato, A.~Garbarz, G.~Giribet and M.~Hassaine,
  Phys.\ Rev.\ D {\bf 80}, 104029 (2009)
  [arXiv:0909.1347 [hep-th]].
  
\bibitem{Sarioglu:2011vz} 
  O.~Sarioglu,
  Phys.\ Rev.\ D {\bf 84}, 127501 (2011)
  [arXiv:1109.4721 [hep-th]].
  
\bibitem{Brynjolfsson:2009ct} 
  E.~J.~Brynjolfsson, U.~H.~Danielsson, L.~Thorlacius and T.~Zingg,
  J.\ Phys.\ A {\bf 43}, 065401 (2010)
  [arXiv:0908.2611 [hep-th]].
  
\bibitem{Balasubramanian:2009rx} 
  K.~Balasubramanian and J.~McGreevy,
  Phys.\ Rev.\ D {\bf 80}, 104039 (2009)
  [arXiv:0909.0263 [hep-th]].
  
\bibitem{Pang:2009pd} 
  D.~-W.~Pang,
  JHEP {\bf 1001}, 116 (2010)
  [arXiv:0911.2777 [hep-th]].
  
  \bibitem{Bertoldi:2009vn} 
  G.~Bertoldi, B.~A.~Burrington and A.~Peet,
  Phys.\ Rev.\ D {\bf 80}, 126003 (2009)
  [arXiv:0905.3183 [hep-th]].
 
\bibitem{Danielsson:2009gi}
  U.~H.~Danielsson and L.~Thorlacius,
  JHEP {\bf 0903} (2009) 070
  [arXiv:0812.5088 [hep-th]].

\bibitem{Dehghani:2011tx}
  M.~H.~Dehghani, R.~B.~Mann and R.~Pourhasan,
  Phys.\ Rev.\ D {\bf 84} (2011) 046002
  [arXiv:1102.0578 [hep-th]].
  
\bibitem{Brenna:2011gp}
  W.~G.~Brenna, M.~H.~Dehghani and R.~B.~Mann,
  Phys.\ Rev.\ D {\bf 84} (2011) 024012
  [arXiv:1101.3476 [hep-th]].
  
\bibitem{Dehghani:2010kd}
  M.~H.~Dehghani and R.~B.~Mann,
  JHEP {\bf 1007} (2010) 019
  [arXiv:1004.4397 [hep-th]].
  
\bibitem{Mann:2009yx}
  R.~B.~Mann,
  JHEP {\bf 0906} (2009) 075
  [arXiv:0905.1136 [hep-th]].


\bibitem{Gubser:2008wv} 
  S.~S.~Gubser and S.~S.~Pufu,
  JHEP {\bf 0811}, 033 (2008)
  [arXiv:0805.2960 [hep-th]].
 
\bibitem{Gubser:2008zu} 
  S.~S.~Gubser,
  Phys.\ Rev.\ Lett.\  {\bf 101}, 191601 (2008)
  [arXiv:0803.3483 [hep-th]].
 
\bibitem{Lu:2013tza} 
  J.~-W.~Lu, Y.~-B.~Wu, P.~Qian, Y.~-Y.~Zhao and X.~Zhang,
  ``Lifshitz Scaling Effects on Holographic Superconductors,''
  arXiv:1311.2699 [hep-th].
 
  
\bibitem{Bartnik:1988am} 
  R.~Bartnik and J.~Mckinnon,
  Phys.\ Rev.\ Lett.\  {\bf 61}, 141 (1988).
\bibitem{Bizon:1990sr} 
  P.~Bizon,
  Phys.\ Rev.\ Lett.\  {\bf 64}, 2844 (1990).
\bibitem{Breitenlohner:1993es} 
  P.~Breitenlohner, P.~Forgacs and D.~Maison,
  Commun.\ Math.\ Phys.\  {\bf 163}, 141 (1994).
\bibitem{Volkov:1998cc}
  M.~S.~Volkov and D.~V.~Gal'tsov,
  Phys.\ Rept.\  {\bf 319} (1999) 1
  [hep-th/9810070].
 
\bibitem{Winstanley:1998sn} 
  E.~Winstanley,
  Class.\ Quant.\ Grav.\  {\bf 16}, 1963 (1999)
  [gr-qc/9812064].
\bibitem{VanderBij:2001ia} 
  J.~J.~Van der Bij and E.~Radu,
  Phys.\ Lett.\ B {\bf 536}, 107 (2002)
  [gr-qc/0107065].
\bibitem{Basler:1986yr}
  M.~Basler,
  J.\ Phys.\ A {\bf 18} (1985) 3087.
  
\bibitem{Basler:1984hw} 
  M.~Basler and A.~Hadicke,
  ``Nonabelian Su(2) Gauge Fields Produced By An Infinite Colored Plane,''
  JENA-N/84/20.
 

\bibitem{Forgacs:1979zs} 
  P.~Forgacs and N.~S.~Manton,
  Commun.\ Math.\ Phys.\  {\bf 72}, 15 (1980).
  
\bibitem{Witten:1976ck} 
  E.~Witten,
  Phys.\ Rev.\ Lett.\  {\bf 38}, 121 (1977).
  
\bibitem{Galtsov:1989ip} 
  D.~V.~Galtsov and A.~A.~Ershov,
  Phys.\ Lett.\ A {\bf 138}, 160 (1989).
 
\bibitem{Mann:2011bt}
  R.~Mann, L.~Pegoraro and M.~Oltean,
  Phys.\ Rev.\ D {\bf 84} (2011) 124047
  [arXiv:1109.5044 [hep-th]].
  
\bibitem{Amado:2011nd} 
  I.~Amado and A.~F.~Faedo,
  JHEP {\bf 1107}, 004 (2011)
  [arXiv:1105.4862 [hep-th]].
  
  
\end{thebibliography}
\end{document}